# The brain as an efficient and robust adaptive learner.

Sophie Denève; Alireza Alemi; Ralph Bourdoukan

Group for Neural Theory, Département d'Etudes Cognitives, Ecole Normale Supérieure, Paris.

## Abstract

Understanding how the brain learns to compute functions reliably, efficiently and robustly with noisy spiking activity is a fundamental challenge in neuroscience. Most sensory and motor tasks can be described as dynamical systems and could presumably be learned by adjusting connection weights in a recurrent biological neural network. However, this is greatly complicated by the credit assignment problem for learning in recurrent network, e.g. the contribution of each connection to the global output error cannot be determined based only on locally accessible quantities to the synapse. Combining tools from adaptive control theory and efficient coding theories, we propose that neural circuits can indeed learn complex dynamic tasks with local synaptic plasticity rules as long as they associate two experimentally established neural mechanisms.  First, they should receive top-down feedbacks driving both their activity and their synaptic plasticity.  Second, inhibitory interneurons should maintain a tight balance between excitation and inhibition in the circuit. The resulting networks could learn arbitrary dynamical systems and produce irregular spike trains as variable as those observed experimentally. Yet, this variability in single neurons may hide an extremely efficient and robust computation at the population level.

## Introduction

The brain is a hugely complex, highly recurrent and nonlinear neural network. This network is surprisingly plastic and sustains our amazing capability for learning from experience and adapting to new situations. It is widely believed that such learning is implemented by synaptic plasticity mechanisms that change synaptic weights as a function of joint pre- and postsynaptic activity (Hebb, 1949). However, most of our neurons are embedded in highly recurrent circuits and several synapses away from sensory receptors or motor effectors. In this situation, it could be generally difficult (if not impossible) to learn global functions like sensory perception, motor control or behavioral tasks based solely on local synaptic plasticity rules. Indeed, while there has been a recent renewed interest for neural networks in machine learning, particularly deep networks (LeCun et al., 2015) and recurrent networks (Sussillo and Abbott, 2009), they use non-local rules such as the "backpropagation" algorithm that are not biologically plausible with the notable exception of feedback alignment methods (Lillicrap et al., 2016). It should be noted that although there have been some proposals for how simple forms of back-propagation could be implemented in a neural structure (Schiess et al., 2016; Urbanczik and Senn, 2014) these are usually limited to two layers. The fundamental reason is one of the credit assignment: in a network where inputs and outputs are several synapses away, and/or where connections are recurrent, there is no trivial way to assign a responsibility for the error made by the entire network to a given synapse. To understand how changing one synaptic weight will affect the network output, one would need to know the synaptic weights of the "downstream" neurons. How, then, can the brain learn complex tasks based on biophysically plausible plasticity mechanisms?

Moreover, neural processing is extremely costly for our metabolism, accounting for about 20% of our total energy consumption. Yet, an extreme level of variability is observed in individual neural responses (Churchland et al., 2010; Tolhurst et al., 1983). The timing and exact number of spikes emitted by cortical neurons is largely unpredictable from trial to trial. Consequently, neural selectivity is usually characterized by a tuning curve (the mean firing rate as a function of the stimulus/task parameter) plus noise. However, increasingly more exhaustive techniques for simultaneous recordings of multiple neurons have revealed that only a small part of the neural response variance is explained by the stimulus or the task. The activity of other nearby neurons is in fact a much better predictor of spike trains (Lin et al., 2015). Computing with such noisy and correlated units could naively appear to be a particularly inefficient and wasteful strategy.

Here, we suggest that these issues can be elegantly resolved by combining two experimentally established properties of neural circuits. First, the maintenance of tight excitatory/inhibitory balance (E-I balance) in a neural system (Brendel et al., 2017; Denève and Machens, 2016; Chalk et al., 2016; Isaacson and Scanziani, 2011; Vogels et al., 2011; Bourdoukan and Denève, 2015; Bourdoukan et al., 2012). Second, a top-down error feedback that modulates both neural circuit activity and synaptic plasticity (Bourdoukan and Denève, 2015). Together, these two basic features of neural circuits can ensure that global functions can be learned with local learning rules, but also that the resulting circuits are highly efficient (e.g. generate minimal neural activity) and extremely robust (i.e. resistant to noise, neural death, etc.) despite the variability in neural responses.

## Balanced networks as efficient autoencoder

Let us start with the simplest problem that a neural population could solve, which is to represent a time-varying input signal, and convey this information to downstream neurons with spikes (Fig 1A). Downstream neurons will perform a synaptic integration of the output spike trains, which can be considered (to a first approximation) as a weighted sum of these output spike trains. The question we ask is how to represent the signal efficiently, i.e. how can the signal be decoded by downstream neurons be as precise as possible, while at the same time requiring the minimal number of output spikes? This may seem to be a hard problem, but it has a simple solution, namely an *autoencoder*: the decoded signal can be directly subtracted from the input by feedback connections, ensuring that the network is only driven by its own coding errors (Fig 1B). Such "error-driven" coding ensures that the representation is very precise: any coding errors will result in an increased feedforward drive, inducing new spikes until the feedforward inputs (i.e. the decoding error) is cancelled. At the same time, this network is maximally efficient: the neurons only respond when it becomes necessary to update the representation. We showed that such network minimizes an objective function corresponding to the coding errors plus a "cost" term penalizing high levels of neural activity (Boerlin et al., 2013). This auto-encoder can also be interpreted as performing a simple form of predictive coding (Rao and Ballard, 1999, 1997).

To obtain a more biologically plausible architecture, we can replace the negative loop by equivalent recurrent connections between neurons. In the fast negative loops, the decoding weights $\mathbf{D}$ are applied to the neural activity to obtain an estimate of the input signal $\hat{\mathbf{s}} = \mathbf{Do}(t)$ [$\mathbf{o}$ is a vector of spike trains]. This estimate is then subtracted from the input, and subsequently fed back into the network via feedforward connections $\mathbf{F}$. The overall effect of the negative loop on neural activity is thus $-\mathbf{FDo}$, which could be implemented by direct recurrent connections (Fig 1C). In this recurrent implementation of the autoencoder, each neuron is driven by its total feedforward input signal, minus the prediction of those input signals by other neurons in the network. In effect, the recurrent connections remove all redundancies between neural spike trains due to their shared input. Interestingly, those lateral connections are also those that achieve the tightest excitatory/inhibitory balance (by "balance" we mean the excitatory and inhibition inputs cancel each other such that a neuron remains very close to the neuronal threshold). The resulting spike trains reproduce the variability of responses observed in cortex, and in particular resemble independent Poisson processes [as indeed predicted in a balanced network (van Vreeswijk and Sompolinsky, 1996)]. We will explain later how this variability can co-exist with a maximally precise coding at the population level.

The last stage to achieve a more plausible network is to replace this recurrent network by an equivalent network of excitatory neurons (pyramidal cells) coupled with inhibitory interneurons (Boerlin et al., 2013; Brendel et al., 2017) (see Fig 1D). While this connection can be set by hand, any network composed of pyramidal cells recurrently connected with interneurons will self-organize into an autoencoder as long as inhibitory connections are trained to cancel excitation as closely as possible. For example, inhibitory connections can be trained to minimize the postsynaptic membrane potential: if the postsynaptic neuron is depolarized at the time of the presynaptic spike, the inhibitory weight is strengthened. On the contrary, if the postsynaptic neuron is hyperpolarized at the time of the presynaptic spikes, the inhibitory weight is weakened [see Fig 1E and (Brendel et al., 2017; Bourdoukan et al., 2012)]. The connections of the resulting network converge to their optimal weights, resulting in a dramatic improvement in the network coding precision (Fig 1F, top panels), a paradoxical increase in single neural variability (middle panels) and tight E-I balance (bottom panels). Endowed with this learning rule, the network can also learn to handle many forms of biophysical limitations such as sparse connections (Maras and Deneve, 2017), noise (Koren and Deneve, 2017), synaptic delays (Chalk et al., 2016; Schwemmer et al., 2015), see also Fig 4C . Because this recurrent inhibition needs to be fast and reliable, the corresponding interneurons would mostly be driven by monosynaptic, fast AMPA synapses from excitatory neurons, targeting the soma and relying on ionic (GABA-A) neurotransmission (e.g. PV interneurons).  There is strong evidence for balanced E-I in cortex (Denève and Machens, 2016).

Note that once the network has learned to track and cancel its own input, purely local biophysical entities such as spikes and membrane potential have acquired functional meanings in terms of the global objective solved by the network (i.e. representing its signal efficiently). For example, a neuron's membrane potential now represents an integration of the global coding error projected onto the feedforward weights $\mathbf{F}$. A spike occurs when this error has reached a threshold, and thus needs to be corrected.  As we will see, the same principle translates to a network that computes more interesting functions of its inputs.

## Adaptive learning of arbitrary dynamical systems

Let us now move to the next stage, and consider how one could learn an efficient implementation of *any function* of the inputs (rather than just an efficient representation of these inputs).  Since both inputs and outputs vary over time, such function could be formalized as a dynamical system, i.e. $\dot{\mathbf{x}} = -\mathbf{x} +$ $f(\mathbf{x}) + \mathbf{s}$, where $\mathbf{s}$ is the input, $\mathbf{x}$ is the time varying state to be represented internally, and $f$ is a function describing the dynamics of $\mathbf{x}$. For the sake of concreteness, we provide a specific example in the framework of sensorimotor control in Table 1. Thus, $\mathbf{x}$ could correspond to the dynamical state of our arm and $\mathbf{s}$ could correspond to an efference copy of the motor commands that were sent to the spinal cord. The function $f$ would represent an internal model of the arm dynamic, able to predict the arm trajectory based solely on its initial position and the sequence of motor commands. Such "forward models" have been identified as key components in building sensorimotor neural controllers (Wolpert and Ghahramani, 2000; Haruno et al., 2001)**.**

Table 1 approximately here

Let us build this forward model step by step, starting from our initial "unfolded" autoencoder (as seen in Fig 1B). First, let us note that we can already interpret the previously described autoencoder model as performing an elementary computation, namely a leaky integration [i.e. $f(\mathbf{x}) = 0$]. For this, we consider the input $\mathbf{s}$ as a leaky derivative of a state variable $\mathbf{x}$, defined by $\dot{\mathbf{x}} + \mathbf{x} = \mathbf{s}$. Meanwhile, we interpret the network output as an estimate of this state variable, i.e. $\hat{\mathbf{s}} = \dot{\hat{\mathbf{x}}} + \hat{\mathbf{x}}$ (Fig 2A). Importantly, this is still *exactly* the same network as in Fig 1B. Since this autoencoder network works by actively cancelling its own input, it automatically enforces that $\dot{\hat{\mathbf{x}}} = -\hat{\mathbf{x}} + \mathbf{s}$. In other words, the autoencoder can be seen as implementing a leaky integration of the input signal $\mathbf{s}$. To recover the state variable $\mathbf{x}$, the spike trains have to be filtered then multiplied by the decoding weights $\mathbf{D}$.

This may appear as a useless exercise at first, but this reinterpretation of the autoencoder becomes interesting once we consider how a non-zero dynamical function $f(\mathbf{x})$ could be implemented. For this, we incorporate an additional loop to the autoencoder (magenta in Fig 2C) whose goal is to approximate $f$ (magenta in Fig 2B). This approximation is injected back into the network as if it was an additional input. This "predictive loop" combines the filtered output spike trains with a set of "slow" connection weights in order to approximate the state dynamic function $f(\hat{\mathbf{x}})$. These connections are called "slow" because the spikes are convolved with an exponential before they are added to the inputs, whereas the negative loop (fast connections) subtracts the spikes directly. This is all that is required if the state dynamics $f$ is linear. If a nonlinear transfer function (e.g. a dendritic or synaptic nonlinearity) is applied to the neural responses before they are combined, any nonlinear function $f$ can be closely approximated (Eliasmith and Anderson, 2004; Eliasmith, 2005; Thalmeier et al., 2016; Abbott et al., 2016; Alemi et al., 2017).

Figure 2 approximately here

With the addition of the predictive loop (magenta in Fig 2B), the effective input to the network becomes $\mathbf{s} - \dot{\hat{\mathbf{x}}} - \hat{\mathbf{x}} + f(\hat{\mathbf{x}})$. Since the negative loop is still in place, the network actively cancels this total input at the short time-scale and ensures that $\dot{\hat{\mathbf{x}}} = -\hat{\mathbf{x}} + f(\hat{\mathbf{x}}) + \mathbf{s}$, and thus that the state estimate automatically follows the desired dynamics. For example, with the appropriate set of slow connection weights, such a network could represent and update an internal estimate of arm position based solely based on the efference copy of the motor commands. We will refer to this network as a "*predictive autoencoder*".

To get a more biologically plausible network, we can fold it again and replace the two sets of feedback loops with two sets of recurrent connections within the network (Fig 2C). We obtain a recurrent network with two types of connections. As previously, fast connections implement the negative loop, and are competitive in nature (similar tuned neurons inhibit each-other). They could be implemented and learned by inhibitory interneurons interconnected with the pyramidal cells (see Fig 1D and the previous section on the autoencoder) and should use fast synapses with ionic receptors (e.g. AMPA/GABA-A) in order to update the postsynaptic neuron's membrane potential as quickly as possible. Slow connections, on the other hand, implement the predictive loop. The input that they provide to postsynaptic cells corresponds to an increase in postsynaptic currents followed by an exponential decay. These could be implemented by slower metabolic channels (e.g. NMDA/GABA-B) and correspond either to direct connections between pyramidal cells, or disynaptic inhibition using another type of interneuron, in the case of negative weights. In contrast to fast connections, slow predictive connections are cooperative in nature: similarly tuned cells tend to excite each other. They will also differ from the fast connections by their learning rules: rather than an unsupervised learning based on balancing excitation and inhibition in the postsynaptic cell, slow connections will be trained based on a feedback from the network's own output errors (see next). When making the network more biologically plausible, the detailed dynamics of synapses turns out to be relatively unimportant. However, one crucially needs connections with different time scales (i.e. filtering properties) to take on (and learn) the strikingly different roles of the negative loop and the predictive loop. Indeed, it is the interplay of fast competition with slower cooperation that allows the network to produce the desired dynamics while remaining efficient and robust (see Fig 4D).

Before we consider how the slow synapses can be learned, it is interesting to compare this framework with the "Neural Engineering Framework" (NEF) (Eliasmith, 2005; Eliasmith and Anderson, 2004) in Fig 2D. The NEF is similar to ours, in that the network injects a prediction of its dynamics back into itself (replaced by recurrent connections in its folded form). The big difference, however, is that NEF does not have a fast corrective loop. It reproduces the dynamics by approximating the derivative of the state dynamics and adding this derivative to the input, but it does not confront this prediction with the current network output in order to correct its own mistakes. This has clear implications in terms of neural coding, learning, efficiency, and robustness, which are different from our framework. In particular, NEF typically exhibits largely regular output spike trains and high firing rates, while the balanced network exhibits irregular spike trains and much lower firing rates. The inherent robustness of the balanced network might also not be present in the NEF implementations.

## Learning the Slow Connections

Let us now imagine that we have access to training examples, i.e. true state trajectories $\mathbf{x}^d(t)$ in response to inputs $\mathbf{s}(t)$, but that the dynamic function $f$ itself is unknown. In the case of a forward model of arm dynamics (table 1), example trajectories could be provided by sensory observations of the arm trajectories as a result of sending specific motor commands to the spinal cord. The goal of learning is then to train the network to reproduce (simulate) similar trajectories as those provided as training examples, but also to generalize to other inputs that were never experienced before. For example, the goal of a forward model of the arm dynamics is to be able to simulate the arm trajectory internally in response to any sets of motor commands, not only those that were used previously.

The actual framework is derived from adaptive control theory. Here, we will only provide some intuitions [For mathematical details of the linear case see (Bourdoukan and Denève, 2015) and for the general nonlinear case see (Sanner and Slotine, 1992; Slotine and Coetsee, 1986; Slotine and Li, 1991)]. Once again, we start from the unfolded network (Fig 3A), with its fast corrective and slow predictive loops. However, the dynamics implemented initially is incorrect. As a result, the network makes large errors: there is a large mismatch between the state estimate $\hat{\mathbf{x}}$ and the desired state $\mathbf{x}^d$.

As we have seen before, learning the slow connections is not trivial due to the credit assignment problem. In particular, changing a single slow connection weight can potentially change the future activity of all neurons in the network and thus have unpredictable consequences for the network output. But what if we could alleviate this non-locality? In particular, if we could force the neural activity to already generate near optimal outputs, we could use the resulting neural activity as a target for learning, without having to worry about unpredictable consequences on the rest of the network. This can be achieved by adding a third "control" loop to the network, this time injecting into the network's input its own supervised error multiplied with a positive gain $K$, i.e. $K(\hat{\mathbf{x}} - \mathbf{x}^d)$ (Fig 3A). If $K$ is large, then the supervised error feedback dominates all the others, and the network represents $\hat{\mathbf{x}} \approx \mathbf{x}^d$. In practice $K$ does not need to be so large and the learning can be proven to converge as long as $K > 0$. In a sense, this is cheating, since all the job is performed by the error feedback (green control loop), not the network connections (magenta predictive loop). Nevertheless, we have now greatly simplified the learning problem. This, as we will see later, will lead to a local plasticity rule for the slow connections. Importantly, as the network learns, its output error becomes smaller and smaller, until it eventually vanishes, i.e. $(\hat{\mathbf{x}} - \mathbf{x}^d) \to \mathbf{0}$. When this occurs, the network will become entirely autonomous (i.e. it does not rely on the "green" control loop anymore).

Figure 3 approximately here

To achieve a more biologically plausible network, we fold the network once more to obtain a recurrent network with slow/fast connections. In addition, each neuron in the network now receives a feedback proportional to the difference between the network's actual output and the desired output (Fig 3B). In the example of the arm forward model, this feedback would correspond to the difference between the perceived and predicted arm position. Such "prediction error" could be provided by feedback connections from sensory areas showing suppression of self-generated sensory signals [i.e. sensory attenuation (Brown et al., 2013)].

The slow connections trained with a local synaptic plasticity rule corresponding to the product of the presynaptic activity and the projection of the error feedback received by the postsynaptic neuron. This could be achieved for example if the slow connections and error feedbacks targeted a similar dendritic region of the postsynaptic neuron (Fig 3C). A simple Hebbian-type learning in the dendrite would lead to a progressive decrease in the error feedback, until it eventually disappears while the feedback connections become silent (Fig 3D left panel). Error feedbacks could also be provided by other means, such as climbing fibers in the cerebellum or backpropagating error potentials in the cortex (Schiess et al., 2016).

Because this framework is derived from adaptive control theory, it inherits its proof of convergence. After learning, the network generates trajectories according to the learned dynamical system f on its own, and can generalize to any new input signals (there is restriction of course, e.g. if the new input is outside the range for which the network was trained). These can be directly inferred from adaptive control theory. This typically requires very short training (e.g. the network only requires enough trajectories to completely determine f, regardless of its size, and thus does not suffer from traditional overfitting). A simple example is provided in Fig 3E. In this case, the network was trained to implement a damped harmonic oscillator.

As for the autoencoder, the learning of the dynamics predictive autoencoder results in biophysical entities such as membrane potential, spikes and synaptic weights, with initially no clear relation to the network objective, acquiring a very specific meaning as a local representation of a global output error defined as the difference between the network output and the learned dynamical system. Because these learning algorithms were derived from top-down principles (efficiency and adaptive learning), they could have widely different neural implementations in different areas/structures/organisms, or even rely on several redundant feedback mechanisms at different spatial and temporal scales. However, given how important it is to learn new tasks quickly, accurately and efficiently, we suspect that what we describe here are very generic principles for neural processing and plasticity in sensory, motor and associative areas. We suspect that similar principle could govern hierarchical learning in the brain, a topic that we intend to explore in the near future (see Discussion). We also envision that very similar mechanisms could implement reinforcement learning (by using the reward prediction errors as a feedback) or even unsupervised learning (by using reconstruction errors of the integrated input itself, e.g. the somatic membrane potential, as the error feedback). In the case of unsupervised learning, the learning rule could be similar to a prospective coding STDP rule proposed recently for prospective coding with spikes (Brea et al., 2016).

## Main properties and testable predictions

We make strong predictions for neural plasticity, variability and robustness, which clearly distinguish this framework from all approaches based on the backpropagation algorithm, and even from its closest relative, the Neural Engineering Framework.

*Prediction 1:* As long as there are significantly more neurons than input signals, the spike trains will have very low firing rates and will be asynchronous and irregular (apparently as noisy as cortical neurons) even if no external noise is injected in the network. This contrasts with the NEF framework that exhibits regular, reproducible and redundant spike trains at high rates (Fig. 2D). This sparseness and variability of spike trains is a consequence of the degeneracy in the neural representations of internal state or external signals.  In Fig 4A, we illustrate this concept with a network composed of only two neurons, with identical decoding weights, representing a one-dimensional time varying signal. The "fast connection" between the two neurons is inhibitory, and ensures that when one neuron spikes, the corresponding decrease in the coding error is taken into account by the other neuron. As a result, when an update in the representation is needed, only one of the two neurons spikes, and resets both neurons. This, in turn, results in a maximally efficient code. Note that individual spike trains look irregular and noisy, even if the global population tracks the signal as accurately as possible. Since the two neurons have the same decoding weights, their firing order is unimportant; any small noise or change in initial conditions will change this order without impacting the representation. In larger networks, this results in asynchronous irregular spikes trains that resemble Poisson processes (see Fig 1F, Fig 3D, Fig 4B,C). The quality of the representations can only be assessed at the population level, not at the single neuron level.  In particular, the precision of the representation scales far better with network size than a rate model with independent Poisson noise [see (Boerlin et al., 2013)].

Figure 4 approximately here

*Prediction 2*: The network is robust to noise, perturbations or destructions of its neurons (Fig 4). Indeed, since the network constantly corrects its own mistakes, even massive perturbations such as suddenly inactivating a large proportion of the neurons (fig 4A,B) will be corrected automatically (without requiring any synaptic plasticity, see fig 4D). This prediction appear compatible with recent reports regarding behavioral and neural robustness after unilateral inactivation of premotor cortex (Li et al., 2016). For a more thorough exploration of the network robustness, see (Barrett et al., 2015). More generally, this robustness implies that the network can still function close to optimal despite biological constraints such as realistic spike generation mechanisms (Schwemmer et al., 2015), delayed synaptic dynamics (Chalk et al., 2016; Koren and Deneve, 2017), sparse connections (Maras and Deneve, 2017), see Fig 4C.

*Prediction 3:* The trained network is very low dimensional (regardless of how large it is in terms of number of neurons/connections). This is true for both the pattern of neural activity it generates, and for the structure of the recurrent connections it learns. The low dimensionality of the fast connection is first enforced by E-I balance. In effect, it shapes neural activity such that $\hat{s} = \mathbf{Dr}$, resulting in a low dimensional structure $-\mathbf{FD}$ for the fast connections. Meanwhile, the low dimensionality of the slow connections is enforced by the error-based, adaptive learning rule. Slow connections will converge to a direct implementation of the learned dynamics whose dimensionality is imposed by $f$, not by the total number of free parameters (connection weights) in the network. In contrast, learning based on temporal back-propagation suffers from severe overfitting, and would probably require far longer training to achieve the same generalization performance.

*Prediction 4:* Error feedbacks (or feedbacks from later processing stages) should simultaneously drive the network and drive synaptic plasticity. In most previous frameworks involving supervised learning in recurrent networks, the output error was used to train the connections, but not to drive the network itself. However, adaptive control works only if the feedback drives the network and pushes its activity in the right direction while synaptic plasticity is taking place (i.e. we need $K > 0$, even if it does not have to be very large). This appears compatible with the fact that feedback connections represent a large portion of connections in the brain, strongly influences neural activity, but also modulates synaptic plasticity (Gilbert and Li, 2013; Zhang et al., 2014). Interestingly, a feedback on the network output (not the error) was introduced in another framework ("FORCE learning") and lead to largely improved performance (Sussillo and Abbott, 2009; Thalmeier et al., 2016).

## Discussion

We have reviewed a new powerful framework, along with its underlying concepts and tools, that allows a spiking neural network to learn arbitrary tasks with maximal efficiency, while respecting basic neural constraints such as local learning rule and robustness [for Dale's law see supplementary materials in (Boerlin et al., 2013)]. The framework makes it possible to implement functional networks that exhibit irregular spike trains very similar to Poisson-like variability of spike trains observed in numerous experimental findings, while being extremely parsimonious in terms of number of neurons and spikes. This should motivate researchers to use this model to explore experimental data for signatures of such spike based (not rate based) coding.

 We believe that application of adaptive control theory, so far relatively unexplored by neuroscientists (at least in recent years) has a bright future for learning functional models of neural circuits. Such purely top-down approaches provide a powerful alternative to bottom up approaches (investigating the influence of a phenomenologically observed learning rule on network dynamics, i.e. (Vogels et al., 2011), or reservoir computing approaches such as FORCE learning (Sussillo and Abbott, 2009). It also raises the enticing possibility that neural representations are much more organized and meaningful than they normally would be in a complex hierarchical and recurrent network like the brain. All that is required is a tight maintenance of E-I balance everywhere in the circuit and feedback connections conveying the difference between the network output and its objective.  Note that this feedback has not one, but two essential roles: 1) its projection drives the network into the right trajectory 2) its projection is directly accessible/available to each neuron and used as the postsynaptic term in the learning of the slow connections. It would be interesting to look for experimental evidence for the simultaneous presence of these two roles in biological neural networks.

The framework is capable of learning highly efficient neural implementations for a variety of behaviorally relevant computations. For example, it can be used to implement working memory (Boerlin and Denève, 2011; Vertechi et al., 2014), dynamical systems for sensorimotor systems (Boerlin et al., 2013), and to implement nonlinear dynamical systems with the help of synaptic nonlinearity (Alemi et al., 2017; Thalmeier et al., 2016).  Furthermore, the framework is generic enough to be, in principle, extended to hierarchical neural circuits in which hidden layer neurons do not have direct access to the global error.

Lastly, it might be worth highlighting the emerging role of the E-I balance and homeostasis as crucial ingredients for neural information processing, computation and learning. Theoretical neuroscience field has long been seeking principles for developing mathematical theories that advance our understanding of mechanisms and functions of neural circuits. Adaptive learning and efficient coding are part of these mathematical theories. E-I balance might a priori look like a phenomena observed in cortical circuits, due to biological constraints at the implementation level (i.e. keep neural activity in check, avoid synchronization…), but without being a computational principle at the algorithmic level (like efficient coding is). What we show here is that E-I balance, efficient coding and adaptive learning are in fact so intimately linked in a spiking network that they might be considered as different sides of the same coin. It would be interesting to explore the implications of disruption of such tight balance. According to the theory outlined here, an outcome of gradually disrupting the tight E-I balance is a degradation of the task and learning performance, which is generally consistent with implication of such disruptions in neurological and psychiatric disorders (Eichler and Meier, 2008; Yizhar et al., 2011; Žiburkus et al., 2013; Denève and Jardri, 2016; Jardri and Deneve, 2015).

## References

Abbott, L.F., DePasquale, B., and Memmesheimer, R.-M. (2016). Building functional networks of spiking model neurons. Nature Publishing Group *19*, 350–355.

Alemi, A., Slotine, J.-J., Machens, C., and Deneve, S. (2017). Learning nonlinear dynamics in spiking networks with local plasticity rules. In Cosyne Abstract, (Salt Lake City, UT), p. 47.

Barrett, D., Deneve, S., and Machens, C. (2015). Optimal compensation for neuron death. bioRxiv 29512.

Boerlin, M., and Denève, S. (2011). Spike-based population coding and working memory. PLoS Comput Biol *7*, e1001080.

Boerlin, M., Machens, C.K., and Denève, S. (2013). Predictive coding of dynamical variables in balanced spiking networks. PLoS Comput Biol *9*, e1003258.

Bourdoukan, R., and Denève, S. (2015). Enforcing balance allows local supervised learning in spiking recurrent networks. In Advances in Neural Information Processing Systems, pp. 982–990.

Bourdoukan, R., Barrett, D., Deneve, S., and Machens, C.K. (2012). Learning optimal spike-based representations. In Advances in Neural Information Processing Systems, pp. 2285–2293.

Brea, J., Gaál, A.T., Urbanczik, R., and Senn, W. (2016). Prospective coding by spiking neurons. PLoS Comput Biol *12*, e1005003.

Brendel, W., Bourdoukan, R., and Vertechi, P. (2017). Learning to represent signals spike by spike. arXiv.org.

Brown, H., Adams, R.A., Parees, I., Edwards, M., and Friston, K. (2013). Active inference, sensory attenuation and illusions. Cognitive Processing *14*, 411–427.

Chalk, M., Gutkin, B., and Deneve, S. (2016). Neural oscillations as a signature of efficient coding in the presence of synaptic delays. eLife.

Churchland, M.M., Yu, B.M., Cunningham, J.P., Sugrue, L.P., Cohen, M.R., Corrado, G.S., Newsome, W.T., Clark, A.M., Hosseini, P., Scott, B.B., et al. (2010). Stimulus onset quenches neural variability: a widespread cortical phenomenon. Nat. Neurosci. *13*, 369–378.

Denève, S., and Jardri, R. (2016). Circular inference: mistaken belief, misplaced trust. Current Opinion in Behavioral Sciences *11*, 40–48.

Denève, S., and Machens, C.K. (2016). Efficient codes and balanced networks. Nature Neuroscience *19*, 375–382.

Eichler, S.A., and Meier, J.C. (2008). EI balance and human diseases-from molecules to networking. Frontiers in Molecular Neuroscience *1*, 2.

Eliasmith, C. (2005). A unified approach to building and controlling spiking attractor networks. Neural Computation *17*, 1276–1314.

Eliasmith, C., and Anderson, C.H. (2004). Neural engineering: Computation, representation, and dynamics in neurobiological systems (MIT press).

Gilbert, C.D., and Li, W. (2013). Top-down influences on visual processing. Nature Reviews Neuroscience *14*, 350–363.

Haruno, M., Wolpert, D.M., and Kawato, M. (2001). Mosaic model for sensorimotor learning and control. Neural Computation.

Hebb, D.O. (1949). The organization of behavior: A neuropsychological approach (John Wiley & Sons).

Isaacson, J.S., and Scanziani, M. (2011). How inhibition shapes cortical activity. Neuron *72*, 231–243.

Jardri, R., and Deneve, S. (2015). Erroneous belief, misplaced confidence. In EUROPEAN PSYCHIATRY, (ELSEVIER FRANCE-EDITIONS SCIENTIFIQUES MEDICALES ELSEVIER 23 RUE LINOIS, 75724 PARIS, FRANCE), pp. S51–S52.

Koren, V., and Deneve, S. (2017). Computational Account of Spontaneous Activity as a Signature of Predictive Coding. PLoS Computational Biology *13*, e1005355.

LeCun, Y., Bengio, Y., and Hinton, G. (2015). Deep learning. Nature *521*, 436–444.

Li, N., Daie, K., Svoboda, K., and Druckmann, S. (2016). Robust neuronal dynamics in premotor cortex during motor planning. Nature *532*, 459–464.

Lillicrap, T.P., Cownden, D., Tweed, D.B., and Akerman, C.J. (2016). Random synaptic feedback weights support error backpropagation for deep learning. Nature Communications *7*, 13276.

Lin, I.-C., Okun, M., Carandini, M., and Harris, K.D. (2015). The nature of shared cortical variability. Neuron *87*, 644–656.

Maras, M., and Deneve, S. (2017). Sparse predictive coding in balanced spiking networks. In Cosyne Abstract, (Salt Lake City, UT), p. 56.

Rao, R., and Ballard, D.H. (1999). Predictive coding in the visual cortex: a functional interpretation of some extra-classical receptive-field effects. Nature Neuroscience.

Rao, R.P., and Ballard, D.H. (1997). Dynamic model of visual recognition predicts neural response properties in the visual cortex. Neural Computation *9*, 721–763.

Sanner, R.M., and Slotine, J.J.E. (1992). Gaussian networks for direct adaptive control. IEEE Transactions on Neural Networks *3*, 837–863.

Schiess, M., Urbanczik, R., and Senn, W. (2016). Somato-dendritic Synaptic Plasticity and Error-backpropagation in Active Dendrites. PLoS Comput Biol *12*, 1–18.

Schwemmer, M.A., Fairhall, A.L., Deneve, S., and Shea-Brown, E.T. (2015). Constructing Precisely Computing Networks with Biophysical Spiking Neurons. The Journal of Neuroscience : The Official Journal of the Society for Neuroscience *35*, 10112–10134.

Slotine, J., and Li, W. (1991). Nonlinear applied control (Prentive Hall).

Slotine, J.J.E., and Coetsee, J.A. (1986). Adaptive sliding controller synthesis for non-linear systems. International Journal of Control *43*, 1631–1651.

Sussillo, D., and Abbott, L.F. (2009). Generating coherent patterns of activity from chaotic neural networks. Neuron *63*, 544–557.

Thalmeier, D., Uhlmann, M., Kappen, H.J., and Memmesheimer, R.-M. (2016). Learning Universal Computations with Spikes. PLoS Comput Biol *12*, 1–29.

Tolhurst, D.J., Movshon, J.A., and Dean, A.F. (1983). The statistical reliability of signals in single neurons in cat and monkey visual cortex. Vision Research *23*, 775–785.

Urbanczik, R., and Senn, W. (2014). Learning by the dendritic prediction of somatic spiking. Neuron *81*, 521–528.

Vertechi, P., Brendel, W., and Machens, C.K. (2014). Unsupervised learning of an efficient short-term memory network. In Advances in Neural Information Processing Systems, pp. 3653–3661.

Vogels, T., Sprekeler, H., Zenke, F., Clopath, C., and Gerstner, W. (2011). Inhibitory plasticity balances excitation and inhibition in sensory pathways and memory networks. Science *334*, 1569–1573.

van Vreeswijk, C., and Sompolinsky, H. (1996). Chaos in neuronal networks with balanced excitatory and inhibitory activity. Science (New York, NY) *274*, 1724–1726.

Wolpert, D.M., and Ghahramani, Z. (2000). Computational principles of movement neuroscience. Nature Neuroscience.

Yizhar, O., Fenno, L.E., Prigge, M., Schneider, F., Davidson, T.J., O�Shea, D.J., Sohal, V.S., Goshen, I., Finkelstein, J., Paz, J.T., et al. (2011). Neocortical excitation/inhibition balance in information processing and social dysfunction. Nature *477*, 171–178.

Zhang, S., Xu, M., Kamigaki, T., Hoang Do, J.P., Chang, W.C., Jenvay, S., Miyamichi, K., Luo, L., and Dan, Y. (2014). Long-range and local circuits for top-down modulation of visual cortex processing. Science (New York, NY) *345*, 660–665.

Žiburkus, J., Cressman, J.R., and Schiff, S.J. (2013). Seizures as imbalanced up states: excitatory and inhibitory conductances during seizure-like events. Journal of Neurophysiology *109*, 1296–1306.

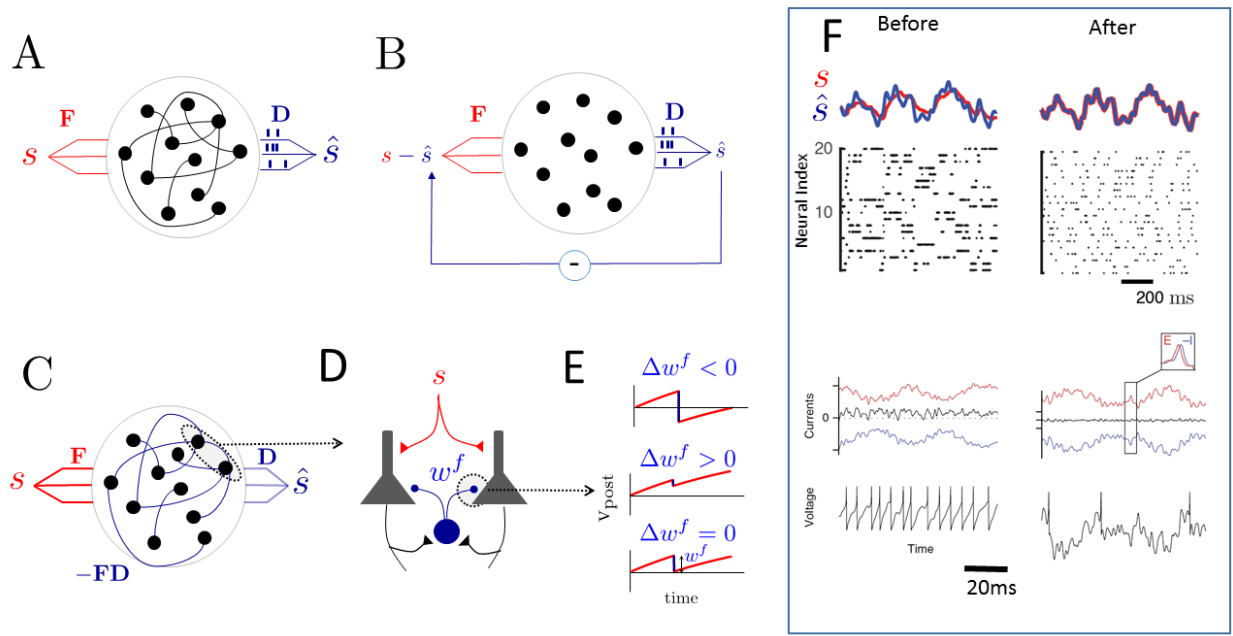

**Figure 1:** Autoencoder network and learning of a tight E-I balance. A. Structure of the recurrent network. Interconnected leaky integrate and fire neurons receive a time-varying input $\mathbf{s}(t)$ with feedforward weights $\mathbf{F}$. Output spikes (vertical bars) are multiplied by the decoding weights $\mathbf{D}$, providing an instantaneous estimate of the input, $\hat{\mathbf{s}}(t)$. B. In this unfolded version of the autoencoder network, the decoded estimate is subtracted from the input. C. Recurrent network equivalent to the unfolded network in panel B. D. Equivalent network of pyramidal cells and interneurons. Interneurons inhibit the pyramidal cells via a fast inhibitory connection with weight $\mathbf{W^f}$. E. Learning rule for training fast connections: the synaptic weight is adjusted to exactly reset the postsynaptic membrane potential (in red) at the time of the pre-synaptic spike (black). It is in fact a Hebbian-inhibitory learning rule with the weight changed according to the product of pre- synaptic spikes and post-synaptic membrane potential (Brendel et al., 2017; Bourdoukan et al., 2012). F. The reconstruction $\hat{\mathbf{s}}$ (in blue) of the input signal $\mathbf{s}$ (in red) is much more precise after learning. Raster plots show the spiking activity in the network before (left panel) and after learning (right panel). [Top and middle panels are adapted from (Bourdoukan and Denève, 2015)]. In the bottom [adapted from (Brendel et al., 2017)], we schematized how excitation dominates in the network before learning, resulting in regular spike trains, while inhibitory current closely tracks excitatory current after learning (with a short time-lag, see inset). Excitatory (red), inhibitory (blue) and total currents are shown above, membrane potentials shown below.

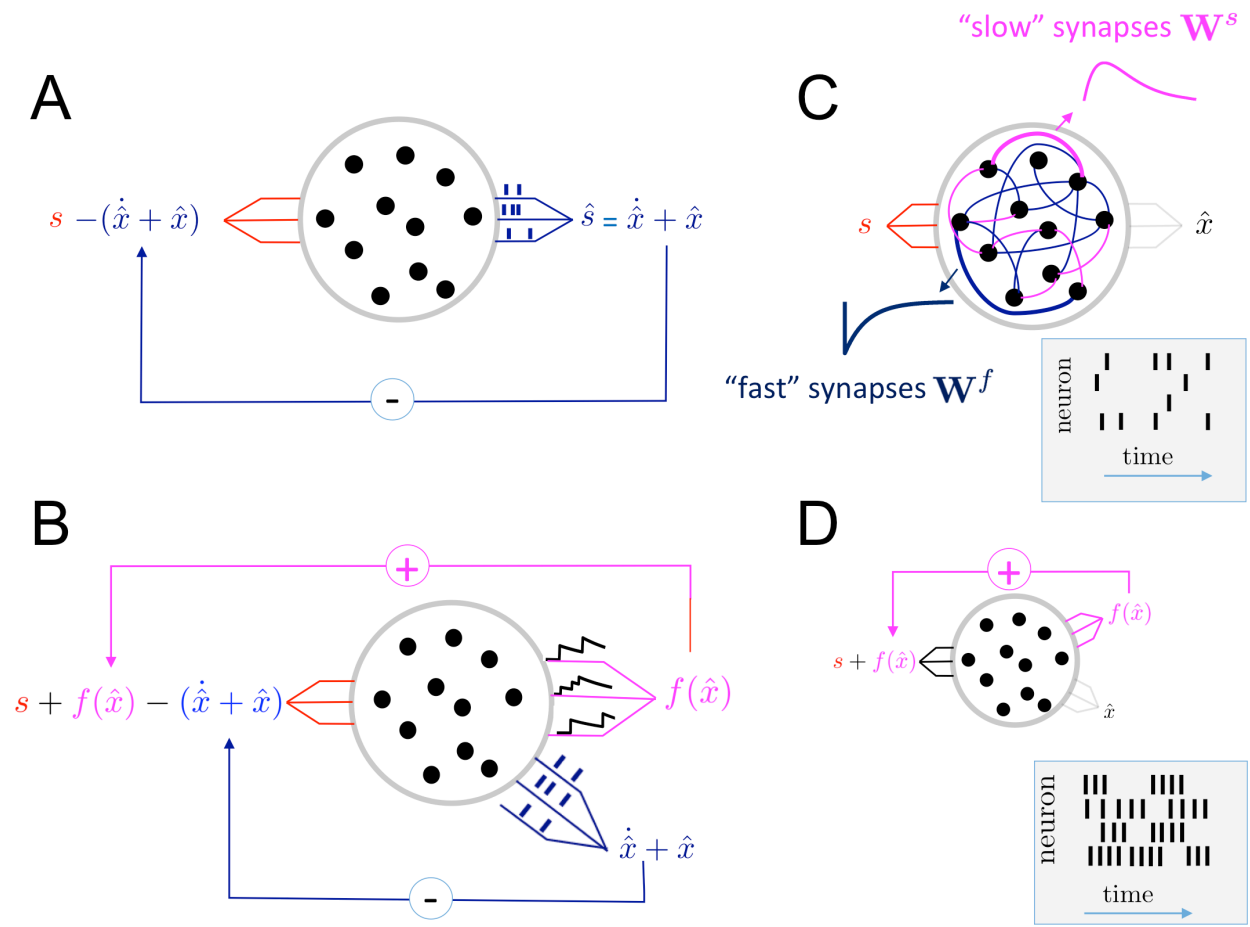

**Figure 2:** Dynamic autoencoder implementing a dynamical system efficiently. A. Alternative interpretation of the unfolded autoencoder in Fig 1B. B. Unfolded network with the addition of a predictive loop approximating the state dynamics (in magenta). Blacks Bars represent the spikes in the fast corrective loop. Black continuous lines represent the filtered spike train in the slower predictive loop. C. Folded network where the fast corrective loop are replaced by fast connections (blue) resulting in fast (exponential kernel) postsynaptic potential (PSP). The slow predictive loop is replaced by another set of connections (magenta) with slower PSP (exponential kernel convolved with itself). D. Schematic representation of an unfolded network in Neural Engineering Framework (Eliasmith and Anderson, 2004).

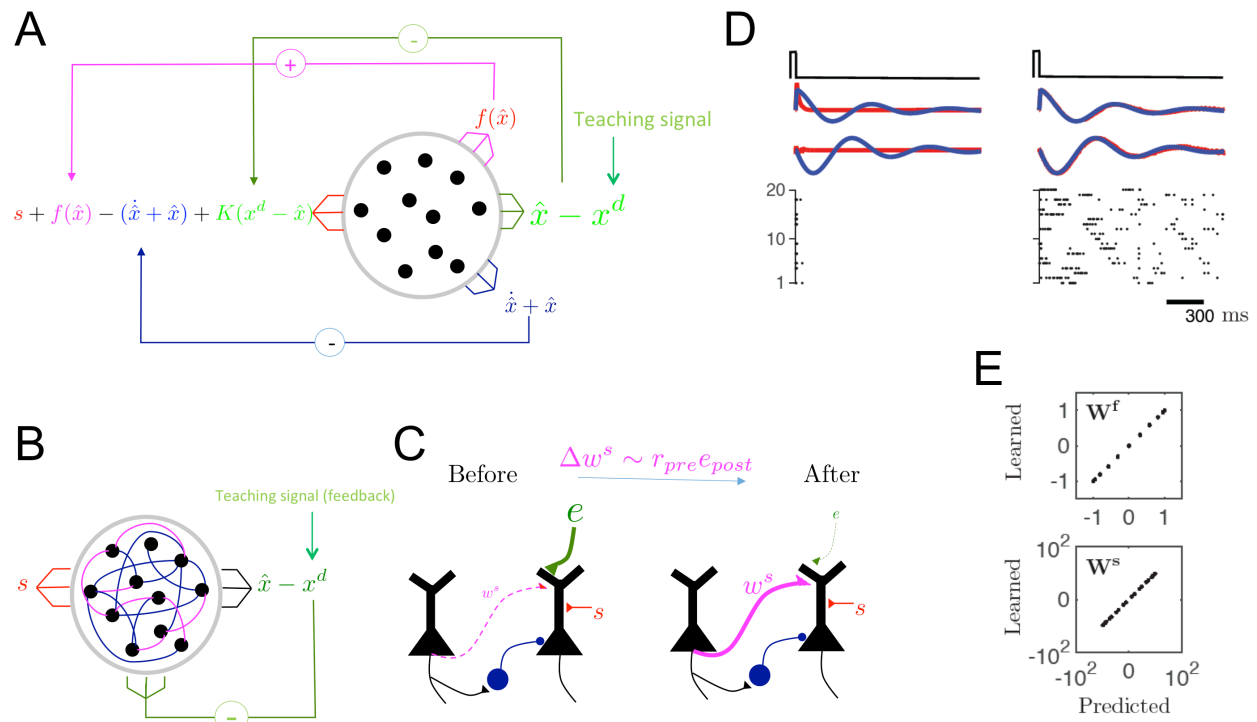

**Figure 3:** Learning the slow connections. A. Unfolded network with the addition of the error feedback (in green). The desired trajectory is provided as a teaching signal and subtracted from the network estimate before being fed-back into the input. B. Equivalent folded network. C. Schema representing the learning of the slow connections. The change in connection weights is proportional to the product between the presynaptic firing rate and the postsynaptic error feedback. Thus, when a presynaptic input is correlated with the error feedback, its weight ($w$) is increased. When a presynaptic input is anti-correlated with the error feedback, its weight ($w$) is decreased. Eventually, the error feedback is cancelled and does not contribute anymore to the learning or neural activity. D. The output of the network (red) and the desired output (blue), before (left) and after (right) shown for learning a damped harmonic oscillator. The black solid line on the top shows the impulse command that drives the network. At the bottom, the spike trains of all network neurons are shown as a raster plot. E. Comparison between the learned weights and their predicted weights based on the optimal closed form solution for the damped harmonic oscillator. Panel D and E are taken from (Bourdoukan and Denève, 2015).

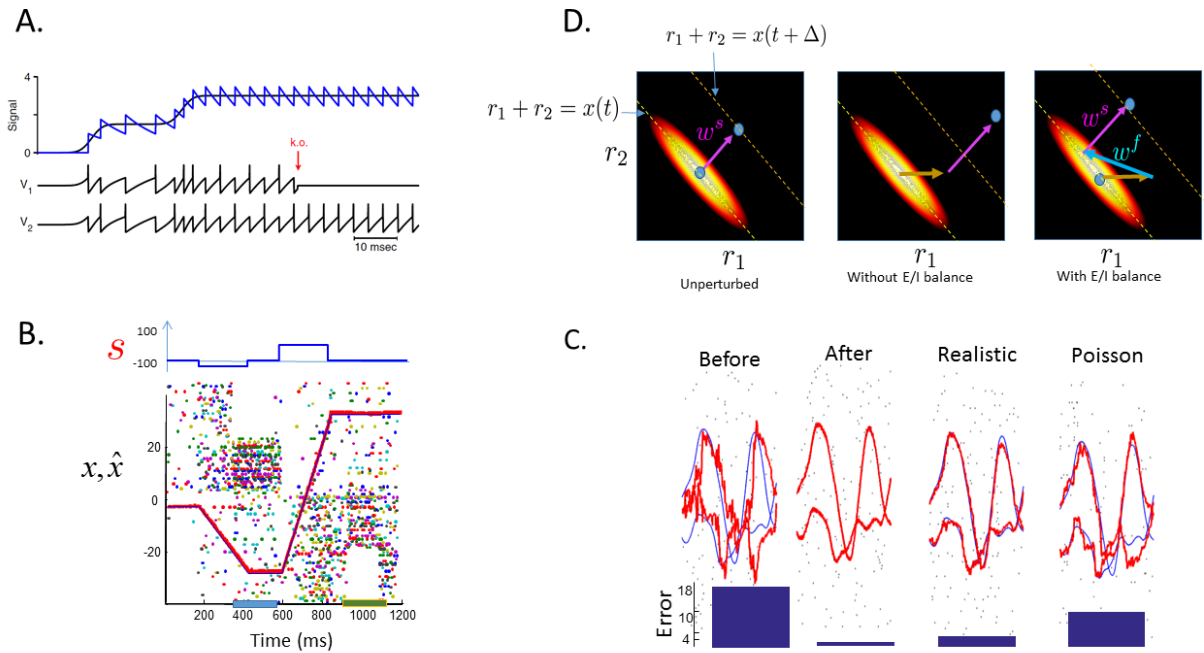

**Figure 4:** Robustness of the predictive auto-encoder. A. A toy example of two neurons reconstructing a continuous input signal (black: filtered input signal, blue: filtered spike train); Bottom panels show the membrane potential and spike trains of the two neurons (adapted from Barrett et al, 2016). B. Robustness of a network of 200 neurons implementing a perfect integrator (adapted from Boerlin et al, 2013). Top panel shows the input signal, dots the spikes, blue and red lines the desired and estimated state, colored boxes indicate time periods during which a quarter of all neurons were inactivated. C. Network performance (100 neurons) before (left panel) and after learning (two middle panels) compared to a population of neurons with independent Poisson noise (right panel). All-to-all connected network with instantaneous synapses '"after" is compared to a "realistic" network with 50% connections of the connections missing, AMPA-like synaptic dynamics and large amount of noise [for details see (Brendel et al., 2017)]. C. Schema illustrating the interplay and fast and slow synapse to ensure the network robustness and efficiency. Left panel: During a time interval $\Delta$, the represented state moves from $\hat{\mathbf{x}}(t)$ to $\hat{\mathbf{x}}(t + \Delta)$ thanks to the slow connection. Neural activity moves between two lines defined as $\mathbf{Dr}(t) = \hat{\mathbf{x}}(t)$ and $\mathbf{Dr}(t + \Delta) = \hat{\mathbf{x}}(t + \Delta)$. For illustration purposes, the network has only two neurons with $\mathbf{D} = [1,1]$, albeit robustness typically requires more neurons. The elongated blob represents the objective function minimized by the network (reconstruction error + cost of neural activity, lighter colors corresponding to a more efficient code). Middle panel: result of a sudden perturbation of neural activity (orange arrow). With no fast connections, the perturbation propagates to the future state estimates. With fast connections (right panel), the activity is brought back toward the optima so quickly that the slow dynamics (and thus, the state estimate) remains unperturbed.

| Mathematical Symbol | Meaning in the model | Example in motor cortex | Putative physiological Mechanism |
|---|---|---|---|
| $\mathbf{D}$ | Decoder | Readout weights for predicting arm position from neural activity | |
| $\mathbf{F}$ | Feedforward connections | Efference copy input strength on each neuron | AMPA/GABA-A |
| $\mathbf{W}^f, w^f$ | Fast connections | Connections to a from inhibitory interneurons | AMPA/GABA-A |
| $\mathbf{W}^s, w^s$ | Slow connections | Recurrent connections between pyramidal cells | NMDA/GABA-B |
| $\mathbf{x}^d(t)$ | Desired dynamical variables | Actual arm position | Sensory feedback (proprioceptive, visual …) |
| $\hat{\mathbf{x}}(t)$ | Estimated dynamical variables | Internal representation of arm position | |
| $\mathbf{s}(t)$ | Input signal | Efference copy of motor command | |
| $\hat{s}(t)$ | Estimated input | Internal representation of the efference copy | |
| $f(\hat{\mathbf{x}})$ | Estimated state dynamics | Forward model of arm dynamics | |
| $\hat{\mathbf{x}}(t) - \mathbf{x}^d(t)$ | Error feedback | Mismatch between predicted and perceived arm position | Prediction error feedback from sensory areas |
| $\mathbf{o}(t)$ | Spike trains | | |
| $\mathbf{r}(t)$ | Filtered spike trains | | |

**Table 1:** The math symbols used in this paper and their corresponding meanings. An example of learning a forward model in sensorimotor learning and putative physiological mechanisms are provided for the quantities.